\nonstopmode
\documentclass[prc,amsmath,showpacs,twocolumn]{revtex4}
\begin{document}
~\\\begin{flushright}TRIUMF preprint: TRI-PP-05-05\end{flushright}
\title{Projection Operator Formalisms and the Nuclear Shell Model}
\author{B.K.~Jennings}\email{jennings@triumf.ca}
\affiliation{TRIUMF, 4004 Wesbrook Mall, Vancouver, BC, Canada V6T 2A3}

\date{\today}
\begin{abstract}
The shell model solve the nuclear many-body problem in a restricted model space
and takes into account the restricted nature of the space by using effective
interactions and operators. In this paper two different methods for generating
the effective interactions are considered. One is based on a partial solution
of the Schrodinger equation (Bloch-Horowitz or the Feshbach projection
formalism) and other on linear algebra (Lee-Suzuki). The two methods are
derived in a parallel manner so that the difference and similarities become
apparent. The connections with the renormalization group are also pointed out.
\end{abstract} 
\maketitle
\pagebreak

\section{Introduction}

The shell model has been in use for many years. There are two main
variants. The first is due to Bloch and Horowitz\cite{bloch}. This method is
closely related to Feshbach projection operator formalism\cite{feshbach}
frequently used to the justify the optical model. The renormalization group
treatment of the two-body scattering due to Birse {\em et al}\cite{birse} is
also closely connected to this method. The second method is due to Lee and
Suzuki\cite{lees}. It is used in the no-core shell model\cite{noc} and to
derive the nucleon-nucleon effective interaction $V_{{\rm low}\
k}$\cite{vlk,schwenk}. It is the purpose the present paper to put the
differences and similarities between the two methods into sharp contrast by a
parallel derivation of the two methods.

\section{The operator $\omega$ and its uses}

We start with a projection operator $P=P^2$ and its compliment
$Q=1-P$. Following Lee-Suzuki\cite{lees} we define an operator:
\begin{eqnarray}
  Q\omega P |\psi\rangle = Q |\psi\rangle \label{one}
\end{eqnarray}
While this is a common part of the Lee-Suzuki method it is rarely used in the
Bloch-Horowitz method. Here we use it in both methods and as will be seen the
main difference between the two methods is precisely in how $\omega$ is
determined. First we note that only the off-diagonal matrix elements of
$\omega$ are needed. In general $\omega$ is not well defined --- a projection
operator does not have an inverse. We need independent information on $|
\psi\rangle$ for example that it is a solution of the Schrodinger equation or a
member of a well define set. However once $\omega$ is known, we can easily
construct effective interactions and operators. For example, starting with the
Schrodinger equation $H|\psi\rangle=E|\psi\rangle$ we have:
\begin{eqnarray}
 &&H(P+Q\omega P)|\psi\rangle=E|\psi\rangle\label{two}\\
 &&P H(P+Q\omega P)|\psi\rangle=EP|\psi\rangle\label{three}\\
 &&(P+ P\omega^\dag Q) H(P+Q\omega P)|\psi\rangle\nonumber\\
                    && \hspace{2.5cm}=E(P+P\omega^\dag
 Q)|\psi\rangle\label{four}\\ 
                    && \hspace{2.5cm}=E(P+P\omega^\dag Q \omega P)
 |\psi\rangle\label{five} 
\end{eqnarray}
where eq.~\ref{one} has been used to eliminate $Q|\psi\rangle$. The form,
eq.~\ref{three}, is the simplest and is the one universally used in the
Bloch-Horowitz method. It has the advantage that $\omega$ occurs linearly. The
last has the advantage of being symmetric. The states $P|\psi\rangle$ are only
orthogonal with with respect to the weight function. Orthogonal functions can
be obtained by defining
\begin{eqnarray}
|\phi\rangle=\sqrt{P+P\omega^\dag Q \omega P} |\psi\rangle \  .\label{phi}
\end{eqnarray}
 The effective interaction is now given as (see ref.~\cite{barrett}):
\begin{eqnarray}
&&\frac{1}{\sqrt{P+P\omega^\dag Q \omega P}} (P+ P\omega^\dag Q) H\nonumber
 \\&&\times(P+Q\omega P) 
 \frac{1}{\sqrt{P+P\omega^\dag Q \omega P}} |\phi\rangle =
 E|\phi\rangle.\label{six}
\end{eqnarray}
The equations, \ref{three} through \ref{six}, are formally equivalent. Which
one to use is matter of taste and convenience. In particular using
$|\psi\rangle$ has the advantage that as we increase the model space we just
add additional contributions leaving the terms already present unchanged
whereas $\phi$ has to be renormalized at each stage. On the other hand the
$\phi$ have the advantage that they are orthogonal (at least for some choices
or $\omega$) and an hermitian operator is being used.

By construction, eq.~\ref{three} gives an effective interaction in the $P$
space that reproduces the $P$ space components of the wave function from the
full-space calculation. If $P$ commutes with the kinetic energy (or more
generally the unperturbed Hamiltonian) then the half off shell t-matrix is also
reproduced since:
\begin{eqnarray}
T|\psi_{PW}\rangle = P  V |\psi\rangle =  V (P+Q\omega P) |\psi \rangle
\end{eqnarray}
where $|\psi_{PW}\rangle$ is a plane wave. Acting on the left side of the last
equation with $P$ restricts the t-matrix to its low momentum components below
the cut off. So we have $PT|\psi_{PW}\rangle= PV (P+Q\omega P) |\psi\rangle$
and can readily identify the low energy effective potential that reproduces the
half off-shell t-matrix as:
\begin{eqnarray}
V_{\rm eff} =P V (P+Q\omega P)\label{eq_eff}.
\end{eqnarray}
This will be true for either the Bloch-Horowitz or Lee-Suzuki choice of
$\omega$. As we will see shortly, the Bloch-Horowitz method reproduces the
fully off-shell t- or k-matrix leading to the same results as Birse {\em et
al}'s\cite{birse} application of the renormalization group.  On the other hand,
in the derivation of the effective interaction, $V_{{\rm low}\ k}$, the authors
only require the half-off shell t-matrix to be reproduced. That approach is
equivalent\cite{vlk} to the Lee-Suzuki method.

It is equally easy to generate the effective operator for a given initial
operator $\Theta$:
\begin{eqnarray}
\langle \psi | \Theta |\psi \rangle = \langle \psi | (P+P\omega^\dag Q)
\Theta  
          (P + Q\omega P) |\psi\rangle\label{oeff}\\ = \langle \phi
         |\frac{1}{\sqrt{P+P\omega^\dag Q \omega P}} (P+P\omega^\dag Q) \Theta
         \nonumber\\
         (P + Q\omega P)\frac{1}{\sqrt{P+P\omega^\dag Q \omega P}} |\phi\rangle
\end{eqnarray}
As these examples illustrate $\omega$ is the key ingredient. Once it is
known, the effective quantities can be calculated easily and the formal
connections become clear. The great advantage of the present method using
$\omega$ is that many of the derivations become very simple.

\section{The Bloch-Horowitz Method}

The first method for solving for $\omega$ is act on eq.~\ref{two} with $Q$,
use eq.~\ref{one} and solve for $\omega$. This gives:
\begin{eqnarray}
Q\omega P|\psi\rangle = \frac{1}{E_\psi
  -QHQ-i\epsilon}QHP|\psi\rangle\label{eq_bbh} 
\end{eqnarray}
There is one aspect of this equation, that thought of as an operator equation,
is quite odd (although so common that one hardly notices it anymore). Namely
the right hand side depends explicitly on the energy of the state --- the
energy has been given a $\psi$ subscript to emphasize this point. The equation
is not a general operator relation but is only valid for a limited set of
matrix elements --- those with the given eigen function of $H$ on the right
hand side. This is, presumably, the source of the starting energy problem: What
energy do we use once we take this equation out of this narrow context? In
spite of all that, if $|\psi\rangle$ is an eigen function of $H$ then
eq.~\ref{eq_bbh} must be satisfied. Even the alternate form for $\omega$
derived in the next section must obey this equation.

Eq.~\ref{eq_bbh} is all that is needed to derive the effective interaction from
eq.~\ref{three}.  Inserting eq.~\ref{eq_bbh} into eq.~\ref{three} we obtain:
\begin{eqnarray}
\left( PHP +PHQ\frac{1}{E_\psi -QHQ-i\epsilon}QHP\right) P|\psi\rangle\nonumber
\\ =
EP\psi\rangle \label{bh}
\end{eqnarray} 
This equation can be immediately recognized as the Bloch-Horowitz shell-model
equation or the Feshbach optical model equation. A bit of a miracle has
occurred here since eq.~\ref{three} is asymmetric with $\omega$ appearing only
on the right hand side while eq.~\ref{bh} is symmetric. Nothing is gained by
going to eq.~\ref{four} or \ref{five}. The effective interaction is not
hermitian due to the $i\epsilon$.  Since $\omega$ is energy dependent both the
effective interaction and the effective operators are energy dependent. For
off-diagonal matrix elements different energies and hence $\omega$'s are needed
on the left and right hand sides of eq.~\ref{oeff}. The energy dependence of
the effective interaction is also a major contributor to making the wave
functions non-orthogonal. The effective interaction and operators can be used
for {\em any} state with a non-zero overlap with the model space. This is in
contrast to the Lee-Suzuki method discussed in the next section.

The effective interaction, eq.~\ref{bh}, can be derived equally well starting
with the Lippmann-Schwinger equation $T=V+VG_0T$. Defining $T_P=PTP$ and
$T_Q=QTP$ and eliminating $T_Q$ gives the equation:
\begin{eqnarray}
T_P=\left( PHP +PHQ\frac{1}{E_\psi -QHQ-i\epsilon}QHP\right)\nonumber\\ + 
\left( PHP +PHQ\frac{1}{E_\psi -QHQ-i\epsilon}QHP\right)\nonumber\\\times
 G_0 T_P\nonumber
\end{eqnarray}
Thus the Bloch-Horowitz effective interaction reproduces the low momentum
components of the fully off-shell t-matrix (or equivalently the k-matrix).
For two-body scattering, it reproduces the results of Birse {\em et
al}\cite{birse}.

\section{The Lee-Suzuki Method}

The second method\cite{lees,lee} of obtaining $\omega$ is a bit more subtle. We
divide the Hilbert space into two parts $|k\rangle$ and $|k'\rangle$. For the
sake of definiteness take the states to be eigenfunctions of the full
Hamiltonian. Consider an orthonormal set of states, $|\alpha_P\rangle$, that
span the $P$ space. If the states $P|k\rangle$ form a complete basis (not
necessarily orthonormal) in the $P$ space then we can write any one of the
$|\alpha_P\rangle$ states as: $ |\alpha_P\rangle= \sum_k P|k\rangle
a^{\alpha_P}_k $.  We now act on this equation with $\langle \alpha_Q| Q\omega
P$ where $\langle \alpha_Q|$ is an arbitrary state in the $Q$ space. Using
eq.~\ref{one} we have:
\begin{eqnarray}
 \langle \alpha_Q| Q\omega P | \alpha_P\rangle &=& \sum_k \langle \alpha_Q |
  Q\omega P |k \rangle a_k^{\alpha_P}\nonumber\\&=& \sum_k \langle \alpha_Q |k \rangle
  a_k^{\alpha_P}
\end{eqnarray}
Since the $|\alpha_P\rangle$ are an orthonormal set the expansion coefficients
$ a_k^{\alpha_P}$ can be determined by inverting the equation:
\begin{eqnarray}
\langle \alpha'_P | \alpha_P\rangle &=& \delta_{\alpha'_P \alpha_P}= \sum_k
\langle \alpha'_P| k \rangle a_k^{\alpha_P}\ . \label{inver}
\end{eqnarray}
This is the defining equation of the Lee-Suzuki method. Once this equation is
solved we have $\omega$ and can calculate the effective interactions and
operators. However the inverse of $\langle \alpha'_P| k \rangle $ must exist.
If not you have to choose a different set of $|k\rangle$'s or modify the
projection operator $P$. The existence of the inverse implies that there are
the same number of states in the $P$ space and the set $|k\rangle$.

The oddity of the Lee-Suzuki method is that $\omega$ depends not only on $P$
but also on the states chosen to be in the set $|k\rangle$. For example, even
if you are only interested in the lowest state, the $\omega$ acting on it will
change if the set $|k\rangle$ is modified to include the 101st state rather
than the 100th.

Note the dramatic differences compared to the previous method. Much to the
surprise of people use to working with the Bloch-Horowitz or Feshbach method the
effective operators in this method are energy independent. Also this method can
work for no more states than the number of states in the $P$ space. This is
also in distinct contrast to the Bloch-Horowitz method.

Since the $V_{{\rm low}\ k}$ obtained in the renormalization group method is
equivalent to the Lee-Suzuki method you might expect that a matrix inversion
like that defined in eq.~\ref{inver} would also play an important role in the
renormalization group derivation. Indeed this is the case, see for example
eq.~4 and 5 in ref.~\cite{schwenk}. The renormalization group $V_{{\rm low}\
k}$ is given by eq.~\ref{eq_eff} with current choice of $\omega$.

\section{Unitary Transformations}

It is well known that the Lee-Suzuki method corresponds to a unitary
transformation\cite{lee} and eq.~\ref{six} does suggests a unitary
transformation with:
\begin{eqnarray}
U=\frac{1}{\sqrt{1+P\omega^\dag Q \omega P}} (P+ P\omega^\dag Q)\ .\label{uni}
\end{eqnarray}
However, $U$ is not a unitary transformation since $UU^\dag=P$ rather than
one. What $U$ does is take a state in the $|k\rangle$ space and transform it
into the $P$ space while $U^\dag$ does the inverse, namely maps a state in the
$P$ space into a state in the $|k\rangle$ space. The operator, $U$, acting on a
state in the $|k'\rangle$ space gives zero. A full unitary operator would have
to map states in $|k'\rangle$ into the $Q$ space. An operator that does this
is:
\begin{eqnarray}
{\cal U} &=&\frac{1}{\sqrt{1+P\omega^\dag Q \omega P}} (P+ P\omega^\dag Q)
          \nonumber\\&&+
           \frac{1}{\sqrt{1+Q\omega P \omega^\dag Q}} (Q - Q \omega P)\\
   &=&\frac{ 1 - (Q\omega
  P- P\omega^\dag Q) }{\sqrt{1+P \omega^\dag Q \omega P + Q \omega P
  \omega^\dag Q}}\ .
\end{eqnarray}
Note $(Q\omega P- P\omega^\dag Q)^2 = -(P \omega^\dag Q \omega P + Q \omega P
\omega^\dag Q) $. When acting on a state in the $|k\rangle$ space this reduces
identically to $U$ since $Q-Q\omega P$ acting on state in $|k\rangle$ gives
zero. Thus if one is only interested in states in the $|k\rangle$ space then
the only difference between the two operators is aesthetics and ink usage. The
full form does indicate that what the Lee-Suzuki method is doing, either
explicitly or implicitly (depending on the formulation) is constructing a
unitary transformation that block diagonalizes the the Hamiltonian.

For a state in the set $|k\rangle$, eq~\ref{uni} can be simplified. Using
eq.~\ref{one} and the definition of $\phi$ (eq.~\ref{phi}) gives $U|k\rangle =
\sqrt{1-P\omega^\dag Q \omega P}P|k\rangle= |\phi_k\rangle$. Using the
completeness of the $|\phi_k \rangle$ in the $P$ space $|\phi_k\rangle$ can be
expanded as $|\phi_k\rangle = \sum_l P|l\rangle \langle k|P|l\rangle^{-1/2}$.
In this form the $|\phi_k\rangle$ are manifestly orthonormal. The wave
functions $P|k\rangle$ span the $P$ space but are not orthonormal. The
$|\phi_k\rangle$ have been orthogonalized in a way that treats all the states
in the $|k\rangle$ space symmetrically.

In the Bloch-Horowitz method it also possible to construct a unitary
transformation however it is much less useful since $\omega$ depends on the
energy of the state of under consideration and the unitary transformation would
be different for each state.

\section{Conclusions}

The important quantity in determining effective interactions or operators is
$\omega$. A projection operator loses information. It is the job of $\omega$ to
restore the lost information. The parallel derivation of the Bloch-Horowitz and
Lee-Suzuki methods shows that the essential difference between them is how the
lost information is recovered or equivalently how $\omega$ is determined.  The
crucial step in the Bloch-Horowitz method is a partial solution of the
Schrodinger equation while in the Lee-Suzuki method the key step is a matrix
inversion.

While the main equations (\ref{three} through \ref{phi}) in the two methods are
similar there are significant differences due to the different methods of
restoring the lost information. The Bloch-Horowitz method deals with one state
at a time and has energy dependent effective interactions and operators.  It
can describe any state with a non-zero overlap with the model space. The
Lee-Suzuki method deals with a set of wave functions, a set with the same
dimension as the $P$ space. It can describe only those states in that set. The
effective interactions and operators are energy independent.
 
Another method for using a restricted space is the renormalization group.  The
crucial equation in comparing to the renormalization group is
eq.~\ref{three}. It shows you how to construct a model-space effective operator
that gives the same model-space components of the wave function as the
calculation in the full space. Either the Bloch-Horowitz or Lee-Suzuki methods
can be used. The first reproduces the fully off-shell t-matrix and is thus
equivalent to the procedure of Birse {\em et al}. The second only reproduces
the half-off shell t-matrix and is equivalent to $V_{{\rm low} K}$.

\section{Acknowledgments}

The author was motived by a seminar by B.~Barrett at TRIUMF and benefited
greatly from discussions during and following a seminar at the Institute for
Nuclear Theory at the University of Washington. Useful discussion where held
with B.~Barrett, W.~Haxton, and A.~Schwenk.  The Natural Sciences and
Engineering Research Council of Canada is thanked for financial support.


\begin{thebibliography}{99}
\bibitem{bloch}C.~Bloch and J.~Horowitz, Nucl.~Phys. 8, 91 (1958).
\bibitem{feshbach}H.~Feshbach, Ann. Phys. (N.Y.) 5, 357(1958); 19, 287 (1962);
B.Block and H.~Feshbach, ibid. 23 47 (1963)
\bibitem{birse}M.C.~Birse, J.A.~McGovern and K.G.~Richardson, Phys.~Lett. B464
  (1999) 169. 
\bibitem{lees}S.Y. Lee and K. Suzuki, Phys. Lett. B91 (1980) 173;
K. Suzuki and S.Y. Lee, Prog. Theor. Phys. 64 (1980) 2091.
\bibitem{noc} P. Navratil, B. R. Barrett, Phys.~Rev. C54 (1996) 298.
\bibitem{vlk} S.K. Bogner, T.T.S. Kuo, A. Schwenk, D.R. Entem, and R. Machleidt
Phys.~Lett. B576 265 (2003).
\bibitem{schwenk}S.K. Bogner, A. Schwenk, T.T.S. Kuo, and G.E. Brown,
nuc1-th/0111042
\bibitem{barrett} P.~Navratil, G.P.~Kamuntavicius, and B.R.~Barrett,
Phys. Rev. C61 (2000) 044001.
b\bibitem{lee}K.~Suzuki, Prog.~Theor.~Phys.  68, 1999 (1982); K.~Suzuki
and
R.~Okamoto, Prog.~Theor.~Phys. 92, 1045 (1994).
\end{thebibliography}
\end{document}